\documentclass{iau} 
\usepackage{graphicx}

\title[The wide-field PISCeS survey] 
{Resolving the extended stellar halos of nearby galaxies: \\ the wide-field PISCeS survey
\footnote{This paper includes data gathered with the 6.5 meter
 Magellan Telescopes located at Las Campanas Observatory, Chile.}
}

\author[D. Crnojevi\'c et al. ]   
{D. Crnojevi\'c$^1$, D. J. Sand$^1$, N. Caldwell$^2$,
  P. Guhathakurta$^3$, B. McLeod$^2$, A. Seth$^4$, J. D. Simon$^5$,
  J. Strader$^6$, E. Toloba$^{1, 3}$}

\affiliation{$^1$Texas Tech University, Physics Department, Box 41051,
  Lubbock, TX 79409-1051, USA 
\\ email: {denija.crnojevic@ttu.edu}\\[\affilskip]
$^2$Harvard-Smithsonian Center for Astrophysics, Cambridge, MA 02138, USA \\[\affilskip]
$^3$UCO/Lick Observatory, University of California, Santa Cruz, 1156 High Street, Santa Cruz, CA 95064, USA \\[\affilskip]
$^4$Department of Physics and Astronomy, University of Utah, Salt Lake City, UT 84112, USA \\[\affilskip]
$^5$Observatories of the Carnegie Institution for Science, 813 Santa Barbara Street, Pasadena, CA 91101, USA \\[\affilskip]
$^6$Michigan State University, Department of Physics and Astronomy, East Lansing, MI 48824, USA \\[\affilskip]
}

\pubyear{2015}
\volume{317}  
\jname{The General Assembly of Galaxy Halos: Structure, Origin and Evolution}
\editors{A. Bragaglia, M. Arnaboldi, M. Rejkuba, \& D. Romano, eds.}
\begin{document}

\maketitle

\begin{abstract}
In the wide-field Panoramic Imaging Survey of Centaurus and
Sculptor (PISCeS), we investigate the resolved stellar halos of two
nearby galaxies (the elliptical Centaurus A and the spiral Sculptor,
D~$\sim3.7$~Mpc) out to a projected galactocentric radius of 150~kpc with
Magellan/Megacam. The survey has led to the discovery of $\sim$20 faint
satellites to date, plus prominent streams and substructures in two environments
that are substantially different from the Local Group, i.e. the
Centaurus A group dominated by an elliptical and the loose Sculptor
group of galaxies. These discoveries clearly attest to the importance of past
and ongoing accretion processes in shaping the 
halos of these nearby galaxies, and provide the first census 
of their satellite systems down to an unprecedented $M_V<-8$. 
The detailed characterization of the stellar content, shape and
gradients in the extended halos of Sculptor, Centaurus A, and their
dwarf satellites provides key constraints on theoretical models of
galaxy formation and evolution.
\end{abstract}

\firstsection 


\section{Introduction}

The past decade has witnessed the advent of wide-field
instrumentation, which has quickly led to the discovery of a wealth
of stellar streams and faint satellites in our own Galaxy's halo
(\cite[Ibata et al. 2001]{ibata01}, \cite[Belokurov et
al. 2006]{belokurov06}). Our nearest massive neighbor, M31,
has been the next to be systematically surveyed out to large
galactocentric radii, uncovering a similarly rich amount of
substructures and satellites (\cite[McConnachie et
al. 2009]{mcconnachie09}). The widely accepted 
$\Lambda$~CDM model for hierarchical structure assembly does indeed
predict that the remnants of past/ongoing accretion and interaction events 
should populate the outskirts of galaxy halos, testifying their 
evolutionary history. Simulations also show a significant
halo-to-halo scatter in the properties of galaxy halos, due to a wide
variety in their assembly histories. However, the physical processes that regulate star
formation and galaxy evolution (e.g. supernova
feedback, reionization, environmental effects, etc.) remain poorly
understood. While the detection of substructures in the
outer halos of virtually all galaxies observed in great depth 
(\cite[Tal et al. 2009]{tal09}, \cite[Martinez-Delgado et
al. 2010]{martinez10}, \cite[Atkinson et al. 2013]{atkinson13},
\cite[Duc et al. 2015]{duc15}) agree well
with theoretical predictions, to date they can only qualitatively confirm
this picture: the intrinsic faintness of these features
($\mu_V\gtrsim28$~mag/arcsec$^2$) poses a challange to
their detailed characterization beyond the Local Group (LG).
To make things worse, there are obvious discrepancies between the predicted
number and baryonic content properties of the smallest galaxies in our
own LG (the ``missing satellite'' and ``too big to fail'' problems, e.g.,
\cite[Moore et al. 1999]{moore99}, \cite[Boylan-Kolchin et al. 2012]{boylan12}).
The faint end of the satellite luminosity function has recently started to be explored
beyond the LG (e.g., M81,
\cite[Chiboucas et al. 2013]{chiboucas13}; M101, \cite[Merritt et
al. 2014]{merritt14}), and yet far fewer galaxies with
$M_V\gtrsim-12$ are observed than predicted by simulations.

In order to test and put quantitative constraints on theoretical 
predictions, it is imperative to observe and characterize 
a larger sample of galaxies, with a range of morphologies and living
in different environments: this has been the motivation for our
Panoramic Imaging Survey of Centaurus and Sculptor (PISCeS), which
we introduce in the next section.


\section{The PISCeS survey}

The PISCeS survey targets two nearby galaxies ($D\sim3.7$~Mpc):
Centaurus~A (Cen~A, or NGC~5128), the closest elliptical to us and the dominant galaxy of
a dense group of galaxies, and Sculptor (NGC~253), a spiral located in
a much looser and elongated group. Sculptor has a mass comparable 
to our own Milky Way (MW; \cite[Karachentsev et
al. 2005]{kara05}), while Cen~A is slightly more massive
(\cite[Woodley et al. 2007]{woodley07}), and due to their
proximity they can be resolved into individual stars, thus allowing
for a detailed comparison with the extant surveys of the MW, M31
and M81.

We use the optical Megacam imager at the 6.5-m Magellan II Clay
telescope (\cite[McLeod et al. 2015]{mcleod15}):
with a field-of-view of $24\times24$~arcmin$^2$ and a binned pixel
scale of $0.16"$, this is an ideal instrument to deliver a deep, wide-field
survey of our target galaxies. With the final goal of reaching projected
galactocentric radii of $\sim150$~kpc (or $\sim16$~deg$^2$) in the
halos of Cen~A and Sculptor, since 2010 we have covered $\sim13$~deg$^2$ 
for each of them in the $g$ and $r$ filters (with $6\times300$~s exposures).
The survey is expected to be completed in 2016.

The standard image reduction and processing is performed by the
Smithsonian Astrophysical Observatory Telescope Data Center, and
subsequently resolved point spread function photometry is obtained
from the stacked images with the DAOPHOT and ALLFRAME packages
(\cite[Stetson 1987, 1994]{stetson87, stetson94}). Standard stars 
from SDSS are observed during clear nights to calibrate the data, and 
adjacent pointings are designed to overlap at the edges to ensure a good calibration under
non photometric conditions as well. 
Finally, we perform artificial star tests in order to assess
photometric uncertainties and incompleteness for each individual
pointing (this is crucial for all such surveys given the
non-uniformity in observing conditions from pointing to pointing).

\begin{figure}
\begin{center}
\includegraphics[width=3.5in]{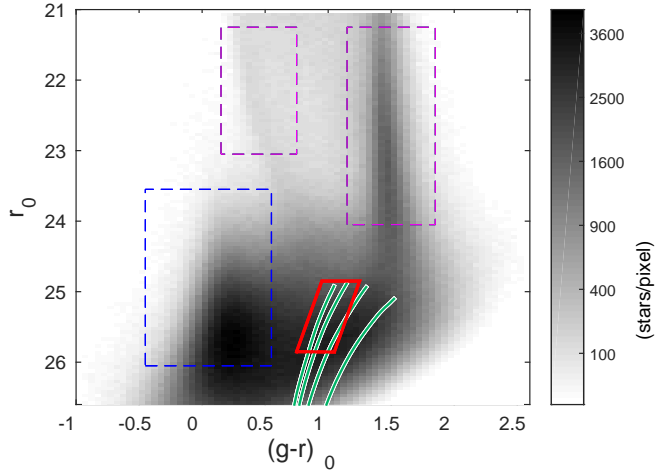} 
 \caption{Dereddened Hess color-magnitude diagram ($0.05\times0.05$~mag
   bins) for the Cen~A PISCeS stellar catalogue to date. Dartmouth
   isochrones (\cite[Dotter et al. 2008]{dotter08}) of 12~Gyr 
with a range of metallicities ([Fe/H]$=-2.0$ to $-1.0$
in 0.5~dex steps) are shifted to Cen~A's distance and overlaid. The
RGB selection box is shown in red, while the contaminants' sequences
are approximately delineated by dashed boxes (blue for unresolved
galaxies at $(g-r)_0\sim0.1$, magenta for MW halo and disk stars, at
$(g-r)_0\sim0.5$ and $(g-r)_0\sim1.5$, respectively).
}
   \label{fig:cmd}
\end{center}
\end{figure}

The deep, resolved Magellan/Megacam images allow us to obtain
color-magnitude diagrams (CMDs) such as the one shown in
Fig.~\ref{fig:cmd} (for Cen~A's survey to date). 
Our survey reaches limiting magnitudes of $\sim$26.5--27~mag in $r$-band, 
which corresponds to $\sim$1.5--2~mag below the tip of the 
red giant branch (TRGB). A red giant branch (RGB) population at the
distance of the target galaxy is clearly identified, as underlined by
the old isochrones with a range of metallicities, and is clearly
separated from the contaminants' sequences (see Fig.~\ref{fig:cmd}).

\section{Preliminary results}

\begin{figure}
\begin{center}
\includegraphics[width=5in]{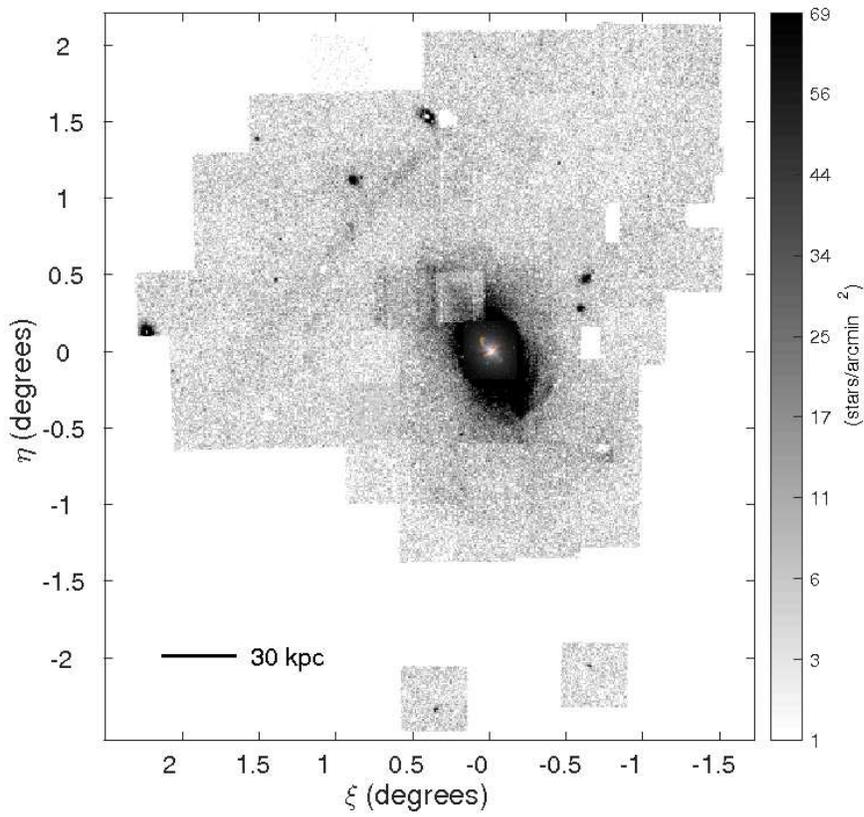} 
\includegraphics[width=3.5in,angle=270]{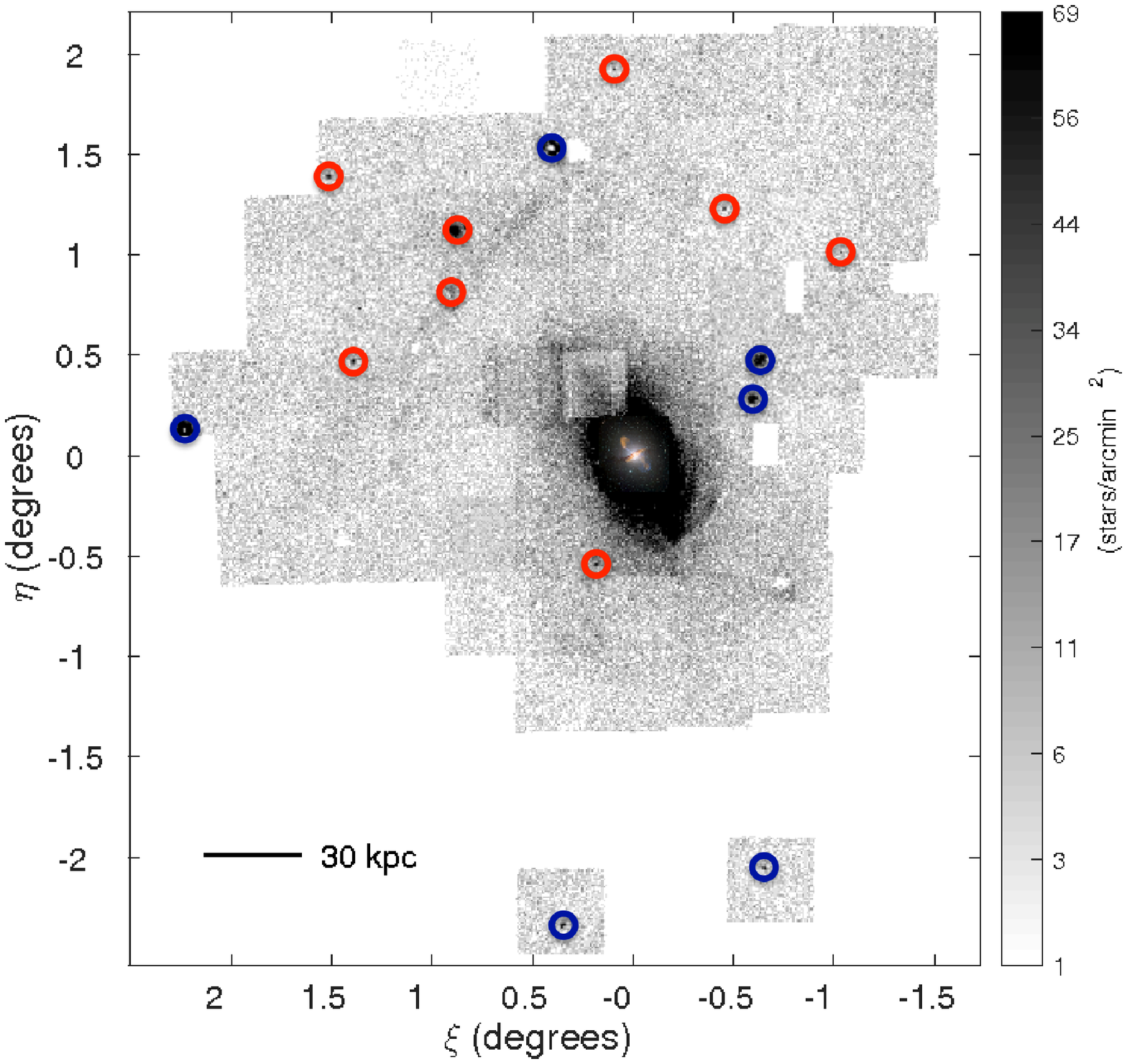} 
 \caption{Stellar density map of metal-poor RGB stars (selected from the box in
   Fig.~\ref{fig:cmd}) for the halo of Cen~A surveyed by PISCeS to
   date. Standard coordinates are centered on Cen~A (N is up and E is
   left), and the central regions of the galaxy are replaced by a color image
(credit: http://www.eso.org/public/images/eso0903a/); the star-count
map in this region suffers from incompleteness due to high stellar crowding.
The density scale as well as the physical scale are shown. In the lower
panel, we indicate the position of the previously known (blue circles) and
newly discovered satellites (red circles).
}
  \label{fig:map}
\end{center}
\end{figure}

To illustrate the potential of our survey, 
we present the metal-poor RGB stellar density map of Cen~A 
in Fig.~\ref{fig:map}, which includes the PISCeS pointings obtained to
date (see Fig.~\ref{fig:cmd} for the RGB selection box). The central
regions of Cen~A clearly show a variety of resolved features including
shells and streams, which highlight the high degree of interaction recently
experienced by this elliptical.

\subsection{Newly discovered satellites}

We use both visual inspection and the detection of overdensities in
the resolved RGB stellar density maps in
order to look for previously unknown faint satellites of Cen~A and
Sculptor. The discoveries of the first robust candidate satellites for
both hosts are reported in \cite{crnojevic14} and \cite{sand14}. In total, to date we
have identified 13 candidate Cen~A satellites and 4 candidates for
Sculptor, down to absolute magnitudes of $M_V\sim-8.0$.
Resolved populations are clearly significantly more powerful than integrated
light alone, given the otherwise prohibitively low surface brightness we are
able to reach (down to $\mu_V\sim32$~mag/arcsec$^2$). Nevertheless,
some of our candidate satellites appear as surface brightness
enhancements coupled with only a few resolved stars, and they may
represent satellites even fainter than our detection limit.
The newly discovered satellites resemble the properties (half-light
radius, central surface brightness) of LG dwarfs with comparable
luminosities (\cite[Crnojevi\'c et al. 2014]{crnojevic14}, \cite[Sand
et al. 2014]{sand14}, \cite[Crnojevi\'c et al. 2015]{crnojevic15},
\cite[Toloba et al. 2015b]{toloba15b}).

Previously known Cen~A satellites are easily recognizable in the
stellar density map (blue circles in the lower panel of
Fig.~\ref{fig:map}): ESO324-24 at 
$(\xi\sim0.3, \eta\sim1.6)$, NGC5237 at $(2.2,0.1)$, KKs55 at 
$(-0.7,0.3)$, KK197 at $(-0.75,0.5)$, as well as KK203 and KK196 in
the two bottom (S) pointings. In the same map, we circle in red the
nine robust new Cen~A satellites, i.e. those that are firmly detected as RGB
overdensities (we do not plot the four remaining candidates that are
only seen as surface brightness enhancements).
These newly discovered satellites are fainter than the known ones 
in terms of both absolute magnitude and central surface
brightness, see e.g. CenA-MM-Dw1 and CenA-MM-Dw2 at $(0.9,1.1)$
(\cite[Crnojevi\'c et al. 2014]{crnojevic14}), or even fainter
candidates at $(1.5,1.4)$ (CenA-MM-Dw17),
$(-0.5,1.2)$ (CenA-MM-Dw3), $(0.2,-0.4)$ (CenA-MM-Dw8) and 
$(0.9,0.8)$ (CenA-MM-Dw16). The latter is the 
remnant of a galaxy caught in the midst of tidal disruption, with
stunning tails extending in both directions over 1~deg across the survey 
footprint (from E to NW). We stress that CenA-MM-Dw16 has extreme
properties (very low central surface brightness coupled with large
extent) with respect to LG satellites within the same luminosity range, and
proves how sensitive our strategy is to such elusive objects. A
similar potentially disrupting satellite, Scl-MM-Dw2, has also been
uncovered in the vicinity of Sculptor (\cite[Toloba et al. 2015b]{toloba15b}).


\section{Conclusions and future work}

The PISCeS survey represents the next observational step in the future of
near-field cosmology beyond the LG. PISCeS will charachterize 
in depth the resolved stellar halos of two $\sim$MW-sized galaxies 
with different morphology and
surrounding environments, and their newly discovered faint satellites
and substructures. We have already secured follow-up 
imaging and spectroscopy of the newly discovered candidate satellites,
streams and substructures with a multi-wavelength approach, namely optical
with HST, near-infrared with Gemini/FLAMINGOS2, co-added spectroscopy
(see also \cite[Toloba et al. 2015a]{toloba15a}) with both VLT/VIMOS and
Keck/DEIMOS. Data have already started to be collected and will be in
hand by the beginning of 2017. This strategy will allow us to derive: 
the profile, shape and extent of the targets' smooth halos and
possible gradients in their resolved populations; the orbital
properties of the uncovered streams which will constrain the host
halo's mass; the relative mass contribution from in situ versus
accreted stellar components; the faint end of the satellite luminosity
function down to $M_V\sim-8.0$, as well as the star formation
histories of the individual satellites discovered through our survey.   
Our results will be compared to extant wide- and narrow-field 
surveys of M31 (PAndAS; SPLASH, \cite[Gilbert et al. 2012]{gilbert12})
and of other nearby galaxies (e.g., M81, \cite[Chiboucas et al. 2013]{chiboucas13},
 \cite[Okamoto et al. 2015]{okamoto15}; GHOSTS, \cite[Monachesi et
 al. 2015]{monachesi15}). This will ultimately  
advance the observational census of galaxy outer halos and 
of their inhabitants, thus providing crucial constraints for current
and future theoretical predictions of galaxy evolution. 


\end{document}